\documentclass[aps,prd,onecolumn,amssymb]{revtex4}
\usepackage{graphicx,bm}
\usepackage[english]{babel}
\usepackage{amsmath}
\usepackage{amsfonts}
\begin{document}

\newcommand{\be}{\begin{equation}}
\newcommand{\ee}{\end{equation}}
\newcommand{\bea}{\begin{eqnarray}}
\newcommand{\eea}{\end{eqnarray}}
\newcommand{\ben}{\begin{enumerate}}
\newcommand{\een}{\end{enumerate}}
\newcommand{\nn}{\nonumber \\}
\newcommand{\e}{\mathrm{e}}

\title{Equation-of-state formalism for dark energy models on the brane \\ and the future of brane universes}
\author{Artyom V. Astashenok$^{a,}$\footnote{E-mail: artyom.art@gmail.com},\ Emilio Elizalde$^{b,}$\footnote{E-mail: elizalde@ieec.uab.es, elizalde@math.mit.edu},\
Sergei D. Odintsov$^{b,c,}$\footnote{E-mail: odintsov@ieec.uab.es. Also at Eurasian National Univ., Astana, Kazakhstan and TSPU, Tomsk, Russia}  and
Artyom V. Yurov$^{a,}$\footnote{E-mail: artyom\_yurov@mail.ru}}

\affiliation{$^{a}$Baltic Federal University of I. Kant, Department of Theoretical Physics,
 236041, 14 Nevsky St., Kaliningrad, Russia \\
 $^{b}$Consejo Superior de Investigaciones Cient\'{\i}ficas, ICE/CSIC and IEEC \\
 Campus UAB, Facultat de Ci\`encies, Torre C5-Par-2a pl, 08193 Bellaterra (Barcelona) Spain\\
$^c$Instituci\'{o} Catalana de Recerca i Estudis Avan\c{c}ats (ICREA)}

\begin{abstract}
Brane dark energy cosmologies, leading to various possible evolutions of our universe, are investigated. The discussion shows that while all these models can be made arbitrarily close to the standard $\Lambda$CDM cosmology at present, their future evolutions can be very different, even diverge with time in a number of ways. This includes asymptotic de-Sitter evolution, Little Rip with dissolution of bound structures, and various possible singularities, as the Big Rip, a sudden future singularity (Type II), and Type III and Type IV cases. Specifically, some interesting effects coming from the brane tension are investigated. It is shown, in particular, that the Little Rip occurs faster on the brane model than in usual FRW cosmology. And in the asymptotic de-Sitter regime the influence of the brane tension leads to a deviation of the value of the effective cosmological constant from that corresponding to ordinary dark energy. As a consequence, the value of the inertial force from the accelerating expansion can greatly exceed the corresponding inertial force in ordinary cosmological models.
\end{abstract}


\maketitle

\section{Introduction}

The discovered cosmic acceleration \cite{Riess,Perlmutter} can be interpreted in terms of the so-called dark energy (for recent reviews, see \cite{Dark-6,Cai:2009zp}), which in a way or other exhibits quite strange properties. However, some attempts at keeping General Relativity (GR) and vacuum fluctuations (sort of a Casimir effect) are still being considered, see e.g. \cite{vari1,Book}, and as well known, with the help of an ideal fluid GR can be rewritten, in an equivalent way, as some modified gravity.

The equation of state (EoS) parameter $w_\mathrm{D}$ for dark energy is known to be negative:
\be
w_\mathrm{D}=p_\mathrm{D}/\rho_\mathrm{D}<0\, ,
\ee
being here $\rho_\mathrm{D}$ the dark energy density and $p_\mathrm{D}$  the pressure.
According to astronomical observations, dark energy currently accounts for some
73\% of the total mass-energy of the universe (see, for instance, Ref.~\cite{Kowalski}).Even if these observations consensually favor the standard $\Lambda$CDM cosmology, being actually rigorous the uncertainties in the determination of the EoS dark energy parameter $w$ are
still too large to define which of the three (quite different) cases $w < -1$, $w = -1$,
and $w >-1$ (remaining always in the vicinity of $w = -1$, of course) is actually realized in our universe, the present observational value (within the framework of the phantom energy model with constant $w_{0}$) being $w=-1.04^{+0.09}_{-0.10}$ \cite{PDP,Amman}. From the theoretical point of view it is interesting to consider models with evolving $w_{0}$, too.

If $w<-1$ (the phantom dark energy case \cite{Caldwell})  all four energy conditions are violated.
According to quantum field theory a field of this kind is unstable. But the possibility to have the effective EOS $w_{\rm eff} < -1$ for dark energy in scalar-tensor gravity without ghosts and other instabilities was shown in \cite{Carrol}. Phantom dark energy can lead to a Big Rip future
singularity \cite{Caldwell,Frampton,S,BR,Nojiri}, where the scale factor becomes infinite at a finite  time in the future.
Another possible scenario is a sudden (Type II) singularity \cite{Barrow} where the scale factor is
finite at the Rip time. However, if $w$ tends to $-1$ sufficiently fast, then no finite-time future singularity occurs \cite{Frampton-2,Frampton-3,Astashenok,LR}. Also, within the framework of braneworld models the
possibility to have $w_{\rm eff} < -1$ without future singularities was originally pointed out in \cite{Sahni0}.
In this case the energy
density increases with time or remains constant. If the energy density grows, the disintegration of any bound structure eventually occurs, in a way similar to the case of the Big Rip singularity.

The future evolution of our universe depends on the choice of the EoS satisfied by dark energy:
\be
\label{EoS-0}
p=g(\rho),
\ee
with $g$ a function of the energy density.

The aim of this article is to develop a general method for the construction
of dark energy models on the brane which are compatible with observational data and which
could provide us with various possible scenarios of universe evolution. This question is
analyzed from the viewpoint of the dark energy EoS and, in particular, from its
corresponding description in terms of a scalar field theory. To this end, Sect. II contains
a brief overview of the EoS formalism in FRW cosmology. Sect. III deals with the explicit  generalization of the EoS formalism to the brane. The classification of cosmological models with different future evolutions is there undertaken. Non-singular models are considered in Sect. IV.
It is demonstrated there that some of these models can actually be made indistinguishable from the ordinary $\Lambda$CDM cosmology, thus remaining stable for very long time before the disintegration of bound structures takes place. In Sect. V  cosmologies which develop future ``soft'' singularities are considered. The properties of these models are seen to be actually similar to those of the previous section. A detailed numerical comparison of the predictions of the models with the observational data leads to the conclusion that they are indeed indistinguishable from the $\Lambda$CDM model in describing the universe evolution up to the present epoch. Moreover, as already mentioned, they may remain stable for billions of years before reaching a future singularity. In this sense, any of these models represents an actually viable alternative to $\Lambda$CDM. A key qualitative result of our study is that current observations make it essentially impossible to determine whether or not the universe will end up in a future singularity. The paper finishes with a  section devoted to discussions.

\section{Dark energy in FRW cosmology: brief overview}

For a spatially flat universe, with metric
\be
ds^{2}=-dt^{2}+a^{2}(t)(dx^{2}+dy^{2}+dz^{2}),
\ee
the cosmological equations are
\be
\label{Fried1}
\left(\frac{\dot a}{a}\right)^2 =  \frac{\rho}{3}\, , \quad
\dot{\rho} =  -3\left(\frac{\dot a}{a}\right)(\rho + p)\, .
\ee
where $\rho$ and $p$ are the total energy-density and pressure, respectively,
$a$ is the scale factor, $\dot{}=d/dt$,
and we use natural system units in which $8\pi G=c=1$.

The EoS for dark energy is conveniently rewritten as
\be \label{EoS}
p_{D}=-\rho_{D}-f(\rho_{D}),
\ee
where $f(\rho_{D})$ is a function of the energy-density. The case $f(\rho_{D})>0$ corresponds
to $w<-1$, while the case $f(\rho)<0$, to $w>-1$. For simplicity, we omit the subscript $D$. If dark energy dominates, one can neglect the contributions of dark matter and radiation. Then, from Eqs.~(\ref{Fried1}), one gets the following link between time and $f(\rho_{D})$:
\be
\label{trho}
t = \frac{2}{\sqrt{3}}\int^{x}_{x_{0}} \frac{d x}{f(x)}\, , \quad
x\equiv\sqrt{\rho}.
\ee
For quintessence the energy density decreases with time, $x<x_{0}$, while for phantom energy $x>x_{0}$.
Let us consider possible variants of the behavior of (\ref{trho}) (compare with \cite{Astashenok}).
\ben
\item The integral (\ref{trho}) converges for $\rho\rightarrow\infty$. Therefore, a finite-time singularity occurs: the energy density becomes infinite at a finite time, $t_{f}$. The expression for the scale factor
\be \label{arho}
a = a_{0}\exp\left(\frac{2}{3}\int^{x}_{x_{0}} \frac{x d x}{f(x)}\right)\, ,
\ee
indicates that there are two possibilities, namely:

\noindent
(1a) the scale factor diverges at finite time (a ``Big Rip'');

\noindent
(1b) the scale factor reaches a finite value and a singularity
($\rho\rightarrow\infty$) occurs. It
is of type III, in the notation of Ref. \cite{Nojiri-2}. The key difference between (1a) and (1b) is that the energy density in the second case grows so rapidly with time that the scale factor does never reach the infinite value.
\item The integral (\ref{trho}) diverges for $\rho\rightarrow\infty$ (these models are described in detail in \cite{Frampton-2,Frampton-3}). The energy density grows with time but
not quickly enough for a Big Rip occurrence. According to the terminology in
\cite{Frampton-2} we have a so-called ``Little Rip'': eventually a dissolution
of all bound structures, one after the other, will happen in the future (the first and very simple example of Little Rip was given in \cite{Ruzmaikina}).

These scenarios can be realized only in the case of having a phantom energy. The next two variants are possible both for a phantom energy and for quintessence.
\item The integral (\ref{trho}) diverges when $\rho\rightarrow\rho_{f}$. The dark energy density tends  asymptotically to a constant value (an ``effective cosmological constant'' \cite{Astashenok}). These models constitute natural alternatives to the
$\Lambda$CDM model, which also leads to a non-singular cosmology. Nevertheless, even for the non-singular asymptotically de Sitter universe, the possibility of a dramatic rip which may lead to the disappearance of bound structures in the universe still remains.
\item Another interesting case occurs if $f(x)\rightarrow\pm \infty$ at
$x=x_{f}$, i.e., the dark energy pressure becomes infinite at finite
energy density. The second derivative of the scale factor diverges while the first derivative remains finite. It is interesting to investigate the family of phantom energy models with
such kind of finite-time future singularities.
\een
An equivalent description in terms of a scalar theory can be obtained using the
equations:
\be
\rho=\pm\dot{\phi}^{2}/2+V(\phi), \quad p=\pm\dot{\phi}^{2}/2-V(\phi),
\ee
where $\phi$ is a scalar field with potential $V(\phi)$. The minus sign in front of the kinetic term corresponds to the phantom energy case. For the scalar field and
potential one derives, respectively, the following expressions
\be
\label{phix}
\phi(x)=\phi_{0}\pm\frac{2}{\sqrt{3}}\int_{x_{0}}^{x}\frac{dx}{\sqrt{|f(x)|}}, \quad V(x)=x^{2}+f(x)/2.
\ee
Combining Eqs.~(\ref{phix}) one obtains the potential as a function of the
scalar field.

\section{Dark energy models on the brane: the equation of state formalism}

We now consider a spacetime which is homogeneous and isotropic along three spatial dimensions, being our 4-dimensional universe an infinitesimally thin wall, with constant spatial curvature, embedded in 5-dimensional spacetime \cite{Sahni,Langlois}. In the Gaussian normal coordinate system, for the brane which is located at $y=0$ one gets:
\be
ds^{2}=-n^{2}dt^{2}+a^{2}(t,y)\gamma_{ij}dx^{i}dx^{j}+\epsilon dy^{2},
\ee
where $\gamma_{ij}$ is the maximally 3-dimensional metric, and $\epsilon=1$, if the extra dimension is space-like, and $\epsilon=-1$, if it is time-like.
Let $t$ be the proper time on the brane ($y=0$), then $n(t,0)=1$. Therefore, one gets the FRW metric on the brane
\be
ds^{2}_{|y=0}=-dt^{2}+a^{2}(t,0)\gamma_{ij}dx^{i}dx^{j}.
\ee
The 5-dimensional Einstein equations have the form
\be
R_{AB}-\frac{1}{2}g_{AB}R=\chi^{2}T_{AB}+\Lambda g_{AB},
\ee
where $\Lambda$ is the bulk cosmological constant. The next step is to write the total energy momentum tensor $T_{AB}$ on the brane as
\be
T^{A}_{B}=S^{A}_{B}\delta(y),
\ee
with $S^{A}_{B}=\mbox{diag}(-\rho_{b},p_{b},p_{b},p_{b},0)$, where $\rho_{b}$ and $p_{b}$ are the total brane energy density and pressure, respectively.

One can now calculate the components of the 5-dimensional Einstein tensor which solve Einstein's equations. One of the crucial issues here is to use appropriate junction conditions near $y=0$. These reduce to the following two relations:
\be
\frac{dn}{ndy}_{|y=0+}=\frac{\chi^{2}}{3}\rho_{b}+\frac{\chi^{2}}{2}p_{b},\qquad \frac{da}{ady}_{|y=0+}=-\frac{\chi^{2}}{6}\rho_{b}.
\ee
After some calculations, one  obtains
\be
H^{2}=\epsilon\chi^{4}\frac{\rho_{b}^{2}}{36}+\frac{\Lambda}{6}-\frac{k}{a^{2}}+\frac{C}{a^{4}}.
\ee
This expression is valid on the brane  only. Here $H=\dot{a}(t,0)/a(t,0)$ and $C$ is an
arbitrary integration constant. The energy conservation equation is correct, too,
\be
\dot{\rho_{b}}+3\frac{\dot{a}}{a}(\rho_{b}+p_{b})=0.
\ee
Now, let $\rho_{b}=\rho+\lambda$, where $\lambda$ is the brane tension. For a fine-tuned brane with $\Lambda=\epsilon\lambda^{2}\chi^{4}/6$, we have the following equation (for $k=0$)
\be \label{BREQ-1}
\frac{\dot{a}^{2}}{a^{2}}=\frac{\epsilon\lambda\chi^{4}}{6}\frac{\rho}{3}
\left(1+\frac{\rho}{2\lambda}\right)+\frac{C}{a^{4}}.
\ee
In this paper we consider a single brane model which mimics general relativity (GR) at present but differs from GR at late times.
We set $8\pi G=\epsilon\sigma\chi^4/6$. As is easy to check, two cases arise: $\epsilon=1$ and $\lambda>0$, or $\epsilon=-1$ and $\lambda<0$. For simplicity, we also set $C=0$ (the term with $C$ is usually called ``dark radiation''). In fact, setting $C\neq0$ does not lead
to additional solutions on a radically new basis, in the framework of our approach.  Eq.~(\ref{BREQ-1}) can be simplified to
\be \label{BREQ}
\frac{\dot{a}^{2}}{a^{2}}=\frac{\rho}{3}\left(1+\frac{\rho}{2\lambda}\right).
\ee
One can see that Eq.~(\ref{BREQ}) for $\rho<<|\lambda|$ differs insignificantly from the FRW equation.  The brane model with a positive tension has been discussed in \cite{Liddle,Sami,Sami-2} in the context of  the unification of early and late time acceleration eras.
The braneworld model with a negative tension and a time-like extra dimension can be regarded as being dual to the Randall-Sundrum model \cite{Sahni-2,Randall,Copeland}. Note that, for this model, the Big Bang singularity is absent. And this fact does not depend upon whether or not matter violates the energy conditions \cite{Asht}. This same scenario has also been used to construct cyclic models for the universe \cite{Cyclic}.

The EoS is taken under the form (\ref{EoS}). It is convenient to put
\be
\rho=\left\{\begin{array} {ll}
2\lambda\sinh^{2}\frac{\eta}{2},\quad \lambda>0, \quad \epsilon=1,\\
2|\lambda|\sin^{2}\frac{\eta}{2}, \quad \lambda<0,\quad \epsilon=-1.
\end{array}
\right.
\ee
The parameter $\eta$ varies from $-\infty$ to $\infty$, for $\lambda>0$, and from $-\pi$ to $\pi$, for $\lambda<0$. Then, one can write equations similar to Eqs.~(\ref{trho}) and (\ref{arho})
\be \label{brtrho}
t(\eta)=\sqrt{\frac{2|\lambda|}{3}}\int_{\eta_{0}}^{\eta}\frac{d\eta}{f(\eta)} ,
\ee
\be \label{brarho}
a(\eta)=\left\{\begin{array} {ll}
a_{0}\exp\left(\frac{\lambda}{3}\int_{\eta_{0}}^{\eta}\frac{\sinh\eta d\eta}{f(\eta)}\right),\quad \lambda>0,\quad \epsilon=1,\\
a_{0}\exp\left(\frac{|\lambda|}{3}\int_{\eta_{0}}^{\eta}\frac{\sin\eta d\eta}{f(\eta)}\right),\quad \lambda<0,\quad \epsilon=-1.
\end{array}
\right.
\ee
The situation is very close to the one for the convenient FRW cosmology. In the case of a positive tension the following possibilities can be realized:
\ben
\item  If the integral (\ref{brtrho}) converges while (\ref{arho}) diverges, we have a Big Rip. It is interesting to note that the Big Rip on a brane considered in \cite{Yurov} occurs faster than in the FRW cosmology. To illustrate this remarkable fact, one can rewrite Eq.~(\ref{brtrho}) in the explicit form:
$$
t(x)=\frac{2}{\sqrt{3}}\int_{x_{0}}^{x}\frac{dx}{\left(1+\frac{x^{2}}{2\lambda}\right)^{1/2}f(x)}.
$$
For the simplest EoS with constant EoS parameter $w_{0}=-1-\alpha^{2}$, the function is $f(x)=\alpha^{2}x^{2}$. If $\rho>>\lambda$, then the dark energy density grows with time substantially faster than in  ordinary cosmology ($\lambda\rightarrow\infty$).
\item  If the integrals (\ref{brtrho}) and (\ref{brarho}) diverge when $\eta\rightarrow\infty$, then a  Little Rip occurs. The acceleration of the universe increases with time definitely faster than in the FRW universe, owing to the brane  tension (see the corresponding time equation for the case (1)).
\item  Asymptotic de Sitter expansion is realized if $f\rightarrow 0$ for $\eta\rightarrow\eta_{f}$, and the integral (\ref{brtrho}) diverges.
\item  There is a type III singularity if both integrals converge when $\eta\rightarrow\infty$.
\item  If $f(\eta)\rightarrow\infty$ for $\eta\rightarrow\eta_{f}$, the universe ends its existence in a sudden future singularity.
\een
The case of negative tension allows for the following interesting possibilities:
\ben
\item  Asymptotic de Sitter expansion, if $f(\eta)\rightarrow0$ for $\eta\rightarrow\eta_{f}$.
\item Asymptotic breakdown (i.e. the rate of expansion of universe tends to 0) will occur  if $f(\eta)\rightarrow 0$ for $\eta\rightarrow\pi$.
\item A sudden future singularity, if $f(\eta)\rightarrow\infty$ when $\eta\rightarrow\eta_{f}$.
\een
For the scalar field and potential we get, respectively, the following equations
\be
\phi(\eta)=\phi_{0}\pm\sqrt{\frac{2|\lambda|}{3}}\int_{\eta_{0}}^{\eta}\frac{d\eta}{\sqrt{|f(\eta)|}}, \quad V(\eta)=f(\eta)/2+\left\{\begin{array} {ll}
\lambda(\cosh\eta-1), \quad \lambda>0,\quad \epsilon=1,\\
|\lambda|(\cos\eta-1), \quad \lambda<0,\quad \epsilon=-1.
\end{array}
\right.
\ee
 One should note that dark energy with an EoS such that $f(\rho)\sim\rho^{\gamma}$, with $\gamma\leq2$, leads to a Big Rip on the brane while, in the case of the conventional FRW universe, such dark energy leads to a Little Rip only.

Some additional remark is in order. Any realistic cosmological scenario should take
into account that dark energy is not a single component of the
universal energy. The key test for dark energy models consist in matching the modulus $\mu$ vs redshift $z=a_{0}/a-1$ relation obtained by the SNe project. As is well known,
\be
\mu(z)=\mbox{const}+5\lg D(z).
\ee
The relation for the luminosity distance $D_{L}(z)$ as a function of the redshift, in the FRW cosmology (FC), is
\be \label{DLFC}
D^{FC}_{L}=\frac{c}{H_{0}}(1+z)\int_{0}^{z}
\left[\Omega_{m}(1+z)^{3}+\Omega_{D}\rho(z)/\rho_{0}\right]^{-1/2}d z\, .
\ee
Here, $\Omega_{m}$ is the total fraction of matter density, $\Omega_{D}$ the fraction of dark energy energy density, and $H_{0}$ is the current Hubble parameter. For simplicity, we neglect the contribution of radiation.
For cosmology on the brane (BC), Eq.~(\ref{DLFC}) can be rewritten as
\be \label{DLBC}
D^{BC}_{L}=\frac{c}{H_{0}}(1+z)\int_{0}^{z}
\left\{\left[\Omega_{m}(1+z)^{3}+\Omega_{D}\rho(z)/\rho_{0}\right]\left[1+\delta(\Omega_{m}(1+z)^{3}
+\Omega_{D}\rho(z)/\rho_{0})\right]\right\}^{-1/2}(1+\delta)^{1/2}d z\, ,
\ee
where the convenient parameter $\delta=(\rho_{0}+\rho_{m})/2\lambda$ has been introduced.

\section{Dark energy models on the brane without final singularities}

The following are examples of exact solutions corresponding to the Little Rip case, asymptotic de Sitter regime, and asymptotic breakdown. It will be assumed, for simplicity, that the universe consists of dark energy only.

\subsection{Little Rip case}
The simplest choice of EoS is to set $f(\eta)=\alpha^{2}=\mbox{const}$. Then,
\be
a(\eta)=a_{0}\exp\left(\frac{\lambda}{3\alpha^{2}}\cosh\eta\right), \quad t(\eta)=\sqrt{\frac{2\lambda}{3}}\frac{\eta}{\alpha^{2}},
\ee
that is
\be
a(t)=a_{0}\exp\left[\frac{\lambda}{3\alpha^{2}}\cosh\left(\sqrt{\frac{3}{2\lambda}}
\alpha^{2}t\right)\right].
\ee
The dependence of the phantom energy density on time reads
\be
\rho(t)=2\lambda\sinh^{2}\left(\sqrt{\frac{3}{2\lambda}}\frac{\alpha^{2}}{2}t\right),
\ee
therefore, a Little Rip occurs on the brane exponentially faster than in the usual FRW cosmology (see \cite{Frampton-2,Astashenok}). The scalar field grows linearly with time, as
\be
\phi=\phi_{0}+\alpha t.
\ee
The potential of the scalar field is exponential, as expected in a string theory:
\be
V(\phi)=2\lambda\sinh^{2}\left(\sqrt{\frac{3}{2\lambda}}\frac{\alpha(\phi-\phi_{0})}{2}\right)+\frac{\alpha^{2}}{2}.
\ee
The acceleration of the universe
leads to an inertial force on any mass, $m$, as seen by a gravitational source
separated by a comoving distance $l$, of the form
\be \label{FIN}
F_{in}=ml\frac{\ddot{a}}{a}=ml\left[\frac{\rho}{3}+\frac{f(\rho)}{2}+\frac{\rho^{2}}{6\lambda}
+\frac{f(\rho)\rho}{2\lambda}\right]
\ee
Following \cite{Frampton-3}, it is convenient to define the dimensionless parameter
\be
\bar{F}_{in}=6\frac{\ddot{a}}{a\rho_{0}}=\frac{2\rho+3f(\rho)+
\rho^{2}/\lambda+3f(\rho)\rho/\lambda}{\rho_{0}}\, ,
\ee
where $\rho_{0}$ is the dark energy density at present.

The first two terms in Eq.~(\ref{FIN}) coincide with the inertial force in the FRW cosmology. From (\ref{FIN}) one can conclude that the dissolution of bound structures in a FRW brane universe occurs essentially faster than in the usual FRW cosmology. The inertial force is a nonlinear function of the energy-density, which in our model reads
\be
\rho=2\lambda\sinh^{2}\left(\gamma t+\mbox{Arsh}\frac{\sqrt{\rho_{0}}}{\sqrt{2\lambda}}\right),\quad \gamma=\frac{\sqrt{3}\alpha^{2}}{2\sqrt{2\lambda}}.
\ee
For $t>>\gamma^{-1}$, one can simply take
\be
\rho\approx\frac{\lambda}{2}\exp(\gamma t)
\ee
and, therefore, the dimensionless inertial force grows exponentially for large times, namely
\be
\bar{F}_{in}\approx\frac{\lambda}{4\rho_{0}}\exp(2\gamma t).
\ee

Performing similar calculations for the classical cosmology, one finds that the inertial force increases with time as $t^{2}$, indeed
\be
\bar{F}_{in}\approx\frac{3\alpha^{2}}{\rho_{0}}(\alpha t)^{2}.
\ee
These models can be matched with the latest data from the Supernova Cosmological Project. For illustration,  let us rewrite the energy density as a function of the redshift. We have
\be \label{RHO_Z}
\rho(z)=\rho_{0}\left[1-\frac{3\alpha^{2}}{\rho_{0}}\ln(1+z)\right]
\ee
and, thus, if the value of the dimensionless parameter $\Delta=3\alpha^{2}/\rho_{0}<<1$, the energy density is, with good accuracy, a constant: our model actually mimics a cosmological constant \cite{vari1,Book}. Therefore, it is trivially, in a way, that our dark energy model describes, for $\Delta<<1$ with $\Omega_{D}=0.72$ and $\Omega_{m}=0.28$, the standard FRW cosmology and shares its celebrated matching with the data coming from astronomical observations . For the brane model one should also define the parameter $\delta$ in Eq.~(\ref{DLBC}). Consequently, at least in the domain $\delta<<1$, the model (\ref{RHO_Z}) on the brane is indistinguishable from  $\Lambda$CDM cosmology.

All structures disintegrate when the inertial force (\ref{FIN}), dominated by
dark energy, becomes equal to the forces which keep them bounded. For the system Sun-Earth, for instance, this happens when $\bar{F}_{in}$ reaches a value $\sim 10^{23}$.
Let us choose, for an example, $\Delta=0.01$ and $\delta=0.01$. In this case, the parameter $\alpha=(\Omega_{D}\Delta)^{1/2}H_{0}\approx0.00624$ Gyr$^{-1}$. The time required for the disintegration of our solar system is
$$
t_{dis}\approx 5\times 10^{14}\mbox{Gyr}
$$
in FRW cosmology, but only of
$$
t_{dis}\approx 9.4\times 10^{5}\mbox{Gyr}
$$
in the brane model.
The disintegration time slowly increases with the growth of the brane tension; to wit,  for $\delta=10^{-19}$ the disintegration time becomes closer to the corresponding time in the FRW cosmology, being then $t_{dis}\approx 0.6\times 10^{14}$ Gyr.

Summing up, we have presented above a realistic Little Rip cosmology on the brane coming from a scalar dark energy with an exponential potential, and have proven that the Little Rip regime sets up remarkably faster here than in the convenient FRW cosmology, owing to the presence of the brane tension.

\subsection{Asymptotic de Sitter regime}
We will now consider the situation corresponding to the case of a brane with negative tension. Let us choose
\be
f(\eta)=\alpha^{2}\cos^{2}(\eta),\quad \alpha=\mbox{const},
\ee
which translates into the following EoS
\be
p(\rho)=-\rho-\alpha^{2}(1+\rho/\lambda)^{2}.
\ee
Time, as a function of the parameter $\eta$, reads
\be
t(\eta)=\left(\frac{2|\lambda|}{3}\right)^{1/2}\alpha^{-2}(\tan\eta-\tan\eta_{0})
\ee
and we see that, when $\eta\rightarrow\pi/2$, then $t\rightarrow\infty$ while $\dot{a}\rightarrow\mbox{const}$. That is, the universe expands in a quasi-de Sitter regime.
The scale factor increases with time as
\be
\label{atAs}
a(t)=a_{0}\exp\left\{\frac{\beta^{2}}{2}\left[\left(1+(\tan\eta_{0}+\alpha t/\beta)^{2}\right)^{1/2}-\cos^{-1}\eta_{0}\right]\right\} .
\ee
In Eq.~(\ref{atAs}) the dimensionless parameter $\beta^{2}=2|\lambda|/3\alpha^{2}$ has been introduced. When $t\rightarrow\infty$, the scale factor grows exponentially  with time,
$$
a(t)\sim\exp\left[(|\lambda|/6)^{1/2}t\right].
$$
Therefore, the effective cosmological constant is $\Lambda_{eff}=|\lambda|/2$. The potential of the scalar field develops a complicated shape. The time dependence of the scalar field is found to be
\be
\phi(t)=-\beta\ln\frac{1-\tan\Phi}{1+\tan\Phi},\quad \Phi=\frac{1}{2}\arctan(\tan\eta_{0}\exp(\alpha t/\beta)),
\ee
where integration constants are omitted. The potential
\be
V(\Phi)=\frac{\alpha^{2}}{2}\cos^{2}2\Phi+\lambda(\cos2\Phi-1)
\ee
tends to $|\lambda|$ with growing time.
Note that, in the brane case, the maximal value of the inertial force (for $\lambda>0$) can considerably exceed the inertial force in the FRW cosmology,
for the chosen EoS for dark energy. To illustrate this possibility, we consider a very simple model, with the following EoS
\be \label{EQ_ASDS}
f(\rho)=\alpha^{2}\left(1-\frac{\rho}{\rho_{f}}\right).
\ee
Observe that, for $\rho_{0}<\rho_{f}$, this equation describes phantom energy while, for $\rho_{0}>\rho_{f}$, it corresponds to quintessence. The energy density-redshift relation reads
\be \label{RHO_ASDS}
\rho(z)=\rho_{f}\left[1-\left(1-\delta_{0}\right)(1+z)^{\Delta\delta_{0}}\right],\quad \delta_{0}=\frac{\rho_{0}}{\rho_{f}},\quad \Delta=3\alpha^{2}/\rho_{0}.
\ee
When $t\rightarrow\infty$ the universe expands in a quasi-de Sitter regime. The scale factor grows exponentially with time, as
\be
a(t)\sim\exp\left(\sqrt{\frac{\Lambda_{eff}}{3}}t \right).
\ee
The effective cosmological constant is then $\Lambda_{eff}=\rho_{f}(1+\delta/\delta_{0})$, and one  concludes that $\Lambda_{eff}>>\rho_{f}$, if $\rho_{f}>>\lambda$.
Also in this case $\rho_{f}>>\lambda$, for the dimensionless inertial force at $t\rightarrow\infty$,  one gets
\be
\bar{F}_{in}\rightarrow\frac{2\rho_{f}}{\rho_{0}}\frac{\rho_{f}}{2\lambda}>>\frac{2\rho_{f}}{\rho_{0}},
\ee
the last expression being the inertial force for the FRW cosmology.

We assume, for simplicity, that $\rho_{f}>>\rho_{0}>>\alpha^{2}$. Using Taylor expansion, one can rewrite Eq.~(\ref{RHO_ASDS}) as
\be
\rho(z)\approx\rho_{0}(1-\Delta\ln(1+z)),
\ee
with the same dimensionless parameter $\Delta$ as in Eq.~(\ref{RHO_Z}). Therefore, as in the former case of the model with a Little Rip, the new model with EoS given by (\ref{EQ_ASDS}), for $\delta<<1$, is also in perfect agreement with the most accurate observational data.
Indeed, if we choose $\rho_{0}/\rho_{f}=10^{-13}$ and let $\delta=0.001$, the value of the dimensionless inertial force on the brane is of the order of $10^{23}$ (and thus, sufficient for the disintegration of our solar system) while it is only of some $10^{13}$ in the case of ordinary FRW cosmology.

As a consequence, on the brane a realistic mild phantom scenario (called a pseudo-Rip) is easy to realize. Owing to the brane  tension, the effective ``cosmological constant'' in brane cosmology could be sufficiently larger than for FRW cosmology and, therefore, disintegration of bound structures may take place in the first case even if the second one remains safe yet.

\subsection{Asymptotic breakdown}
Asymptotic breakdown can be realized only for a brane model with negative tension and provided the extra dimension is time-like.
If $\eta\rightarrow\pi$, the Hubble parameter tends to zero i.e., the universe bounces when the dark energy density has reached
a sufficiently large value. Therefore, the expansion rate of the universe decreases with time. There are two possibilities: (i) the scale factor grows, albeit more and more slowly with time, or (ii) the scale factor tends to a constant value. For the first case, consider
\be
f(\eta)=\alpha^{2}\sin^{2}\eta,
\ee
where the EoS in explicit form can be written as
\be
p=-\left(1+\frac{2\alpha^{2}}{|\lambda|}\right)\rho+\frac{\alpha^{2}}{\lambda^{2}}\rho^{2}.
\ee
Then,
$$
t(\eta)=\frac{\beta}{\alpha}(\cot\eta-\cot\eta_{0}),\quad a(\eta)=a_{0}\left(\frac{\cot\frac{\eta_{0}}{2}}{\cot\frac{\eta}{2}}\right)^{\beta^{2}/2}
$$
and, therefore,
\be
a(t)=a_{0}\left[\frac{x+(1+x^{2})^{1/2}}{\cot\frac{\eta_{0}}{2}}\right]^{\beta^{2}/2} ,\quad x=\alpha t/\beta-\cot\eta_{0}.
\ee
The  asymptotic behavior of this expression is power-like when $t\rightarrow\infty$ ($x\rightarrow\infty$), namely
$$
a(t)\sim t^{\beta^{2}/2}.
$$
If one chooses $f(\eta)=\alpha^{2}\sin^{2}\eta$, then the universe expansion will asymptotically stop. After a rather simple calculation, we obtain for the scale factor
$$
a(t)=a_{0}\exp\left[\beta^{2}\arctan\left(\tan\frac{\eta_{0}}{2}\exp(\alpha t/\beta)\right)-\beta^{2}\frac{\eta_{0}}{2}\right].
$$
Therefore, $a\rightarrow a_{0}\exp[\beta^{2}(\pi-\eta_{0})/2]$ when $t\rightarrow\infty$. It is interesting to note that, at some point, the accelerated expansion of the universe turns into deceleration.
As one can easily see from the following expression for the second derivative of the scale factor,
\be
\frac{\ddot{a}}{a}=\frac{9\alpha^{4}\sin\eta}{4\lambda^{2}}(\sin\eta+\cos\eta),
\ee
this instant corresponds to $\eta=3\pi/4$.

\section{Dark energy models on the brane with future ``soft'' singularities}
In this section we study brane dark energy models which lead to Type II or Type III future singularities and compare the corresponding predictions with the ones of the standard FRW cosmology \cite{Astashenok}.

 \subsection{Big Freeze singularity (BFS, Type III)}
Let us consider this case for $\lambda>0$. Take
\be
f(\eta)=\alpha^{2}\cosh^{2}\eta
\ee
which leads, in explicit form, to the following EoS
\be
p=-\rho-\alpha^{2}\left(\frac{\rho}{\lambda}+1\right)^{2},
\ee
a choice which yields a solution exhibiting two BFS: one in the past ($\eta\rightarrow-\infty$) and another in the future ($\eta\rightarrow+\infty$). Time is given by
\be
t(\eta)=\sqrt{\frac{2\lambda}{3}}\frac{\tanh\eta}{\alpha^{2}},
\ee
and the universe begins at $t_{in}=-\frac{1}{\alpha^{2}}\sqrt{\frac{2\lambda}{3}}$ and ends its existence at $t_{f}=\frac{1}{\alpha^{2}}\sqrt{\frac{2\lambda}{3}}$. The expression for the scale factor reads
\be
a(t)=a_{f}\exp\left[-\frac{\lambda}{3\alpha^{2}}\left(1-\frac{3\alpha^{4} t^{2}}{2\lambda}\right)^{1/2}\right],
\ee
where $a_{f}$ is the final value of the scale factor. This dependence shows that the universe will contract in the interval $t_{in}<t<0$ and then it will expand.

Can a model like this be matched with actual observational data? Remarkably, the answer is positive. Indeed, after a simple calculation we obtain, for the dark energy density as a function of the redshift, $z$,
\be
\rho=\rho_{0}\ \frac{1-(1/2+\Omega_{D}\delta)\Delta\ln(1+z)}{1+(1+2\Omega_{D}\delta)
\Delta\Omega_{D}\delta\ln(1+z)}, \quad \Delta=3\alpha^{2}/\rho_{0}.
\ee
Once more, we see that when $\delta<<1$ and $\Delta<<1$, our model precisely mimics $\Lambda$CDM-cosmology, specifically in the range $0<z<1.5$. The time at the final singularity is
\be
t_{f}\approx\frac{1}{\Delta \Omega_{D} H_{0}}\delta^{-1/2}.
\ee
In particular, if $\Delta=0.1$ and $\delta=0.01$, then the time elapsed before the singularity is formed is of some $2\times10^{3}$ Gyr.
Other models with similar properties can be readily constructed in an analogous way.

\subsection{Sudden future singularity (SFS, Type II)}
Let us now discuss realistic models of dark energy
which contain a Type II future singularity for the brane with a negative tension. To this end, we choose the following EoS
\be
f(\eta)=\frac{\beta}{\cos^{2}(\eta-\eta^{*})},\quad 0<\eta^{*}<\pi/2.
\ee
where $\beta$ is a constant, either negative or positive. This EoS describes both phantom energy (when $\beta>0$, $0\leq\eta_{0}<\pi/2$) and quintessence (when $\beta<0$, $\pi/2<\eta_{0}\leq\pi$). The explicit function $f(\rho)$ reads
\begin{equation}\label{EOSSFS}
f(\rho)=\frac{\beta}{(1-\rho/\lambda^{*})^{2}},\quad \lambda^{*}=2|\lambda|\sin^{2}\left(\frac{\pi}{4}+\frac{\eta*}{2}\right)>|\lambda|.
\end{equation}
When $\rho\rightarrow\lambda^{*}$, the pressure diverges and a SFS is formed. The Hubble parameter remains finite but the acceleration of the universe diverges, $\ddot{a}\rightarrow\pm\infty$.

Let's consider the case when $\eta^{*}$ is close to $0$. With good accuracy one can derive the following relation
\be
\rho(z)\approx|\lambda|\left\{1-\left[\left(1-\frac{\rho_{0}}{|\lambda|}\right)^{3}+
\frac{9\beta}{|\lambda|}\ln(z+1)\right]^{1/3}\right\}.
\ee
If $\beta<<\rho_{0}<<|\lambda|$ (phantom energy), this model mimics $\Lambda$CDM cosmology in the range of $z$ corresponding to the SNe observations:
$$
\rho(z)\approx\rho_{0}(1-\Delta\ln(z+1)),\quad \Delta=3\beta/\rho_{0}.
$$
The time remaining until the final singularity is formed can be calculated from Eq.~(\ref{brtrho}). The instant the SFS occurs corresponds to $\eta=\pi/2+\eta^{*}\approx\pi/2$. Therefore, one has
\be
t_{SFS}\approx\left(\frac{2|\lambda|}{3}\right)^{1/2}\frac{1}{\beta}\left(\frac{\pi}{4}-\frac{\eta_{0}}{2}
-\frac{\sin2\eta_{0}}{4}\right),
\ee
and taking into account that $|\delta|<<1$, and correspondingly $\eta_{0}<<\pi/2$,
$$
t_{SFS}\approx\left(\frac{2|\lambda|}{3}\right)^{1/2}\frac{1}{\beta}\frac{\pi}{4}
\approx\frac{\pi}{4}\frac{1}{H_{0}}\frac{1}{|\delta|^{1/2}\Delta\Omega_{D}}.
$$
In particular, if $\delta=-0.01$ and $\Delta=0.1$, the remaining time is $1.5\times10^{3}$ Gyr. Acceleration of the universe turns into deceleration due to the negative tension of the brane. The universe acceleration as a function of the density is (we neglect here matter contribution and assume that dark energy dominates at present)
\be
\frac{\ddot{a}}{{a}}=\frac{\rho}{3}\left(1+\frac{\rho}{2\lambda}\right)
+\frac{2\delta\Delta\Omega_{D}\lambda}{3(1-\rho/\lambda^{*})^{2}}\left(\frac{1}{2}
+\frac{\rho}{2\lambda}\right).
\ee
When dark energy density reaches $|\lambda|$ the second term in this equation vanishes and then changes sign from ``$+$'' to ``$-$". Near the sudden future singularity ($f(\rho)\rightarrow\pm\infty$), the expression for the inertial force (\ref{FIN})  reduces to
\be \label{ASFIN}
\bar{F}_{in}\approx\frac{3f(\rho)+3f(\rho)\rho/\lambda}{\rho_{0}}.
\ee
In our case, $\rho\rightarrow\lambda^{*}$, Eq.~(\ref{ASFIN}) yields
$$
\bar{F}_{in}\approx\frac{3f(\rho)}{\rho_{0}}\left(1-\frac{\rho}{|\lambda|}\right)
=\frac{\Delta}{(1-\rho/\lambda^{*})^{2}}\left(1-\frac{\rho}{|\lambda|}\right)
$$
and hence, due to the negative tension on the brane, we have a so-called ``Big Crush'': at $\rho\rightarrow\lambda^{*}>|\lambda|$ the acceleration of the universe  $\ddot{a}\rightarrow-\infty$, instead of $\ddot{a}\rightarrow\infty$ as for the usual FRW universe.

To finish, we can consider the scalar description of the model under investigation. For the scalar field, as a function of $\eta$, we have
$$
\phi(\eta)=\phi_{f}\sin(\eta-\eta^{*}),\quad \phi_{f}=\left(\frac{2|\lambda|}{3|\beta|}\right)^{1/2},
$$
where the integration constant has been omitted. From here, the potential as a function of the scalar field is given by
\be
V(\phi)=\frac{\beta}{2}\frac{1}{1-\phi^{2}/\phi_{f}^{2}}+\lambda((1-\phi^{2}/\phi_{f}^{2})^{1/2}-1).
\ee
When $\phi$ is small ($\eta\rightarrow 0 $ in the phantom case, and $\eta\rightarrow \pi$ for quintessence) this potential corresponds to a massive scalar field, namely
$$
V(\phi)\approx\frac{|\lambda|}{2\phi_{f}^{2}}\phi^{2},\quad \phi\approx0.
$$
Then, the sudden future singularity is here the singularity of the potential at $\phi=\phi_{f}$. In exactly the same way, comparison of other brane dark energy models, leading to soft singularities in the future with the predictions coming from the ordinary FRW dark energy, can be done.

\section{Conclusion}

Several dark energy models on the brane, with various different scenarios of evolution have been
presented. We must emphasize the remarkable property, shared by all these models, that they are able to mimic, under quite reasonable conditions, the standard $\Lambda$CDM cosmology at the present epoch but exhibit, otherwise, very different behaviors concerning their future evolutions, including asymptotic de-Sitter evolution, Little Rip behavior with eventual dissolution of bound structures, and various different future singularities (as the Big Rip one, Type II and Type III singularities).

The future evolution of the universe is actually determined by the selected EoS for the dark energy component. The presence of a brane tension has been proven to lead to very interesting effects. In particular, owing to its influence the Little Rip occurs faster on the brane than in the usual FRW universe. Also, at a pseudo-Rip the effective value of the vacuum energy does not coincide with the dark energy density and, consequently, the disintegration of all bound structures can actually occur. Furthermore, a negative brane tension changes acceleration into deceleration and leads to a ``Big Crush'' of the universe, its existence ending, in this case, in a sudden future singularity.

These results can be eventually used to discriminate the theoretical dark energy models when confronting them with precise cosmological data. The consequences of all these calculations are to be taken into account quite seriously, since they could actually compromise the degree of predictiveness commonly attributed to the more standard cosmological scenarios.

\subsection*{Acknowledgements.}
The work by AVY has been supported by the ESF, project 4868 ``The cosmological constant as eigenvalue of Sturm-Liouville problem'', and the work by AVA has been supported by the ESF, project 4760 ``Dark energy landscape and vacuum polarization account'', both within the European Network ``New Trends and Applications of the Casimir Effect''. EE's research has been partly supported by MICINN (Spain), contract PR2011-0128. EE and SDO have been supported in part by MICINN (Spain) projects FIS2006-02842 and FIS2010-15640, by the CPAN Consolider Ingenio Project, and by AGAUR (Generalitat de Ca\-ta\-lu\-nya), contract 2009SGR-994. EE's research was partly carried out while on leave at the Department of Physics and Astronomy, Dartmouth College, 6127 Wilder Laboratory, Hanover, NH 03755, USA.

\end{document}